# OPTIMAL CHANNEL ALLOCATION WITH DYNAMIC POWER CONTROL IN CELLULAR NETWORKS

Xin Wu, Arunita Jaekel and Ataul Bari

School of Computer Science, University of Windsor
401 Sunset Avenue, Windsor, ON, N9B 3P4, Canada
{wu11f, arunita, bari1}@uwindsor.ca

## ABSTRACT

*Techniques for channel allocation in cellular networks have been an area of intense research interest for many years. An efficient channel allocation scheme can significantly reduce call-blocking and call-dropping probabilities. Another important issue is to effectively manage the power requirements for communication. An efficient power control strategy leads to reduced power consumption and improved signal quality. In this paper, we present a novel integer linear program (ILP) formulation that jointly optimizes channel allocation and power control for incoming calls, based on the carrier-to-interference ratio (CIR). In our approach we use a hybrid channel assignment scheme, where an incoming call is admitted only if a suitable channel is found such that the CIR of all ongoing calls on that channel, as well as that of the new call, will be above a specified value. Our formulation also guarantees that the overall power requirement for the selected channel will be minimized as much as possible and that no ongoing calls will be dropped as a result of admitting the new call. We have run simulations on a benchmark 49 cell environment with 70 channels to investigate the effect of different parameters such as the desired CIR. The results indicate that our approach leads to significant improvements over existing techniques.*

## 1. INTRODUCTION

Due to the tremendous growth in the demand for mobile communication services, it is important to have a channel assignment scheme that can allocate these limited resources fairly and efficiently. In a cellular network, the geographical area is partitioned into cells where each cell has a base station and a number of mobile terminals (e.g. mobile phone) [1]. The communication between a mobile terminal and the rest of the information network is managed by the base station [1]. For a new incoming call, assigning an appropriate channel for each communication request that arrives in a cell is known as the channel assignment problem. The power control scheme assigns an appropriate power level to the incoming call, with the objective of minimizing total power consumption and suppressing interference. Both the channel assignment and the power control problems, by themselves, are known to be difficult problems [2], [3]. Therefore, many researchers have traditionally considered these two problems separately [2], [4], [5], i.e. first a channel is selected and then a power control algorithm is used to check and update the carrier-to-interference ratio (CIR) value of the channel.

In order to meet the required signal quality specifications, the assigned channel and its power level (for the incoming call) should be selected in a way that ensures that the CIR for the new call and all ongoing calls are maintained above an acceptable level. Existing techniques typically use heuristics to solve this problem, which generally leads to sup-optimal solutions. In this paper we present an integer linear program (ILP) formulation that can be used to *optimally* solve the combined channel assignment and power control problem in wireless cellular networks. Our approach ensures that the CIR requirements are met, not only for the new incoming call but also for all ongoing calls using the same channel. This means that, unlike many previous approaches [6], [7], [8], the *call dropping* probability due to an incoming call is zero. If CIR requirements cannot be met for all ongoing calls, then the new call is simply blocked. Furthermore, our approach assigns an appropriate power level to the incoming call and all ongoing calls using the selected channel such that the overall power consumption is *minimized*. This leads to a significant reduction in blocking probability. We study the performance of our formulation for different values of CIR and also investigate how the blocking probability varies with different ratios of fixed and dynamic channels for each cell. The results demonstrate that our approach clearly outperforms channel assignment schemes that do not address power control issues, as well as existing heuristics that combine channel assignment and power control.

The remainder of the paper is organized as follows. In Section 2, we review the relevant work on channel assignment and power control schemes for cellular networks. In Section 3, we present our ILP formulation. We discuss our experimental results in Section 4 and conclude in Section 5.

## 2. REVIEW

### 2.1. Channel Assignment

The availability of channels or frequency spectrum is very limited, as compared to the exponential growth of mobile terminals, requiring approaches for sharing the channels for efficient assignment and proper management of channel resources [1]. The allocation of the frequencies to mobile terminals and the base stations such that the network's capacity, in terms of number of mobile users, is maximal is known as the channel assignment problem [1]. This is a well-known NP-hard problem [3] and has been widely investigated in the literature.

Generally, channel assignment schemes (without power control) can be divided into three categories; fixed channel-assignment (FCA), dynamic channel-assignment (DCA), and hybrid channel-assignment (HCA). The FCA schemes [9-11] allocate channels permanently to each cell based on predetermined estimated traffic. FCA scheme is simple but it does not adapt to changes in traffic conditions. In a cell, a channel can be assigned to a call using FCA, only if there are free channels available in the predetermined set for this cell. Otherwise, the call might be rejected even though there are many channels available in the network.

In DCA schemes [12-16], the channels are assigned on a call-by-call basis in a dynamic fashion and the entire set of channels is accessible to all the cells. DCA offers more efficient networks, especially under known distribution of the traffic load or when the traffic load varies with time [1]. Since, in DCA, channel assignment takes into account current network conditions, it offers flexibility and traffic adaptability. DCA methods have better performance than fixed channel assignment methods for light to medium traffic load [21].

The Hybrid channel-assignment (HCA) [22, 21, 23, 24] includes features from the FCA and the DCA approaches, and hence, overcomes their drawbacks. Hybrid channel-assignment divides the set of channels into the following two subsets [21],

1. The *Fixed Channels (FC)* set: channels permanently allocated to the cells, and
2. The *Dynamic Channels (DC)* set: channels available to all cells.

The ratio, FC:DC, of the number of channels in FC set and DC set, for a given number of channels is fixed and assigned by the cellular network designer. For example, for a set of 70 channels, the values for ratios FC:DC can be set to: 35:35, 49:21 or 21:49. We note that in case the FC set is NULL, then the HCA problem can be seen as a simple DCA problem.

## 2.2. Power Control

A mobile terminal needs to obtain a channel from the base station for beginning communicating with a base station. A channel consists of a pair of frequencies as follows:

1. The *down-link* frequency: used to communicate from the base station to the mobile terminal, and

2. The *up-link* frequency: used to communicate from the mobile terminal to the base station.

Here, we focus on the down-link transmission only.

The relevant propagation effects are modelled by the link gains as shown in Fig. 1. Let $B_j$ and $M_j$ denote the base station and mobile terminal respectively in cell $j$, and let $g_{ij}$ denote the link gain from the transmitter at $B_j$ (for any channel $l$), to the receiver at $M_i$ using the same channel in cell $i$. The gain $g_{ii}$ corresponds to the desired communication link, whereas $g_{ij}; i \neq j$ corresponds to unwanted co-channel interferences.

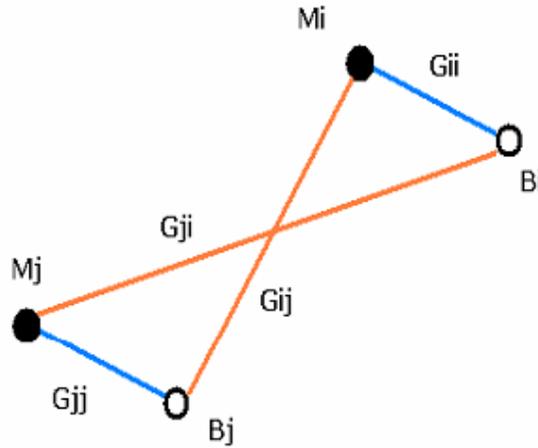

Fig. 1. System Geometry and Link Gains

Let, $p_{jl}$ be the transmitter power on channel $l$ at base station $B_j$. The signal power received at the receiver of $M_i$ on the same channel, from the base station transmitter in cell $j$ is $g_{ij} \cdot p_{jl}$. The desired signal level at the receiver in cell $i$ is equal to $g_{ii} \cdot p_{il}$, while the total interfering signal power from other transmitters (using the same channel $l$ for ongoing calls in their respective cells) to the receiver in cell $i$ is $\Sigma_{j \neq i} g_{ij} \cdot p_{jl}$. As in [4], we use the CIR at $M_i$ (denoted by $\Gamma_i$) as measure of the signal quality at mobile $M_i$.

$$\Gamma_i = \frac{g_{ii} \cdot p_{il}}{\Sigma_{j \neq i} g_{ij} \cdot p_{jl} + \eta_i}, \quad 1 \leq i,j \leq C \qquad (1)$$

where $\eta_i > 0$ is the thermal noise power at mobile $M_i$. The CIR is acceptable if $\Gamma_i$ is above a certain threshold, $\gamma_0$, called the minimum protection ratio. This $\gamma_0$ reflects some minimum quality of service (QoS) that the link must support throughout the transmission in order to operate properly. Hence, for acceptable CIR, have:

$$\frac{g_{ii} \cdot p_{il}}{\Sigma_{j \neq i} g_{ij} \cdot p_{jl} + \eta_i} \geq \gamma_0 \qquad (2)$$

DCA approaches and power control schemes [2], [4], [5], [17], [18], [19] are considered as the most efficient techniques to manage scarce resource available to the system. Many researches demonstrated that a remarkable improvement in terms of spectrum efficiency and network capacity could be gained via integrating two techniques together [19], [20], [26]. Most existing DCA algorithms such as those in [6], [7], [8], do not guarantee that all existing calls will maintain the desired CIR. The works in [4], [5] ensure that existing calls will not be dropped, but do not jointly optimize both channel allocation and power control. Unlike previous techniques, our approach jointly performs both channel allocation and power control and adjusts power levels to guarantee that ongoing calls are not dropped.

In [27], the authors discuss structure of optimal decentralized power control policies for mobile units with discrete power levels. In [28], closed-loop power control and adaptive beam-forming is discussed and the performance of *direct sequence-code division multiple access* cellular systems is studied in urban environment.

## 3. INTEGRATED CHANNEL ASSIGNMENT AND POWER CONTROL

We assume that each base station in a cellular network has a computer to store information about the current state of its cell. The state of the cell includes information about channels in use for ongoing calls, the transmitter power levels for each channel, and associated link gains. The information for each base station is made available to other base stations through a wired network between their computers. This setup is similar to those used in [2], [21].

When a new call arrives in cell k, the computer of the concerned base station is responsible for allocating an unused channel l to the call and determining the transmitter power level to be used. The introduction of the new call may increase the interference levels in neighboring cells. Therefore, it is important that power levels of ongoing calls, using the same channel l allocated to the new call, be adjusted if necessary to ensure that the CIR for ongoing calls do not fall below the specified threshold.

In order to service an incoming call in a cell k at time t, we first search for a channel in the fixed channel set (FC). If no such channel is available from FC then we use the proposed ILP formulation to determine a suitable channel and transmitter power level from the dynamic channel set (DC) for the new incoming call in cell k. The ILP also evaluates if any adjustment is needed to transmitter powers for the same channel in other cells, in order to maintain CIR of ongoing calls. If so, the ILP determines the lowest power levels for the selected channel in all cells, such that no ongoing calls will be dropped. The calculated power levels are then communicated to the respective base station computers in each cell via the wired network. On receiving this information, the base stations update their power levels accordingly. In this paper, we consider only down-link transmission. However, the proposed strategy can be easily used for up-link transmissions as well. Our approach does not require any reassignment of channels for ongoing calls in the network.

Our ILP utilizes the information in the channel allocation matrix (A), defined in the next section, in order to find a suitable channel and appropriate transmitter power levels in each cell. The channel allocation matrix contains information about the channels currently in use in each cell in the network, and is updated every time a channel is allocated or released in the network. Each base station receives a copy of the updated allocation matrix.

### 3.1. Notation used

In the ILP formulation, the following symbols represent the input data:
- $k$: Cell where a call arrives.
- $d_k$: Number of calls in cell $k$ (traffic demand in cell $k$), including the new call.

- $C$: Number of cells in the network.
- $M$: Maximum transmission power level.
- $L$: Number of dynamic channels.
- $B$: Set $\{1, 2, ..., L\}$ of channel numbers for all dynamic channels.
- $B_f$: Subset of $B$, containing the channels currently not in use in cell $k$.
- $d_{i,j}$: Normalized distance between cell $i$ and cell $j$, $1 \leq i; j \leq C$.
- $g_{i,j}$: An element of a $C \times C$ matrix $G$, where each element, $g_{i,j}$, indicates the gain between cells $i$ and $j$.
- $a_{i,l}$: An element of a $C \times L$ allocation matrix $A$, where each element, $a_{i,l}$, is defined as follows: $a_{i,l} = \begin{cases} 1, & \text{if channel } l \text{ is in use in cell } k. \\ 0, & \text{otherwise} \end{cases}$

We also define the following variables:

- $w_l = \begin{cases} 1, & \text{if channel } l \text{ is selected for the new call in cell } k, \forall l \in B_f. \\ 0, & \text{otherwise} \end{cases}$
- $p_{i,l}$: A continuous variable indicating the transmitter power level of channel $l \in B_f$ in cell $i$, $1 \leq i \leq C$.

### 3.2. ILP Formulation

We now present our ILP formulation that allocates a free channel to a new call such that the total power is minimized. Using the notation given above, we formulate the ILP as follows:

Objective function:

$$\text{Minimize} \quad \sum_{i=1}^{C} \sum_{l \in B_f} p_{i,l} \quad (3)$$

Subject to:

1. Only one channel per call.

$$\sum_{l \in B_f} w_l = 1 \quad (4)$$

2. No power on idle channels.

$$p_{i,l} \leq a_{i,l} \cdot M, \forall i, i \neq k, \forall l \in B_f \quad (5)$$

$$p_{i,l} \leq w_l \cdot M, \forall i,, \forall l \in B_f \quad (6)$$

3. CIR requirements must be satisfied.

$$g_{i,i} \cdot p_{i,l} \geq \gamma_0 \cdot \sum_{i \neq j} g_{i,j} \cdot p_{j,l} + \gamma_0 \cdot \eta_i \cdot w_l \quad (7)$$

Constraint (7) must be satisfied for all $l$ such that $l \in B_f$ and for all $i$ such that $i = k$, or $a_{i,l} = 1$.

The objective function specified in (3) selects an available channel $l$ for the new call in cell $k$, such that the total transmission power required to maintain an acceptable CIR for the new call as well as all existing calls on channel $l$ is minimized.

Constraint (4) enforces that, from the pool of available dynamic channels ($B_f$), exactly one channel that is currently not in use in cell $k$ is allocated for a new call. Constraint (5) states that in all cells $i$ ($i \neq k$), there is non-zero transmission power on channel $l$ if and only if there is an ongoing call using that channel. Constraint (6) states that power levels for all channels other than the channel $l$ selected for the new call (i.e. the channel for which $w_l = 1$) is set to 0. We note that this does not imply that all transmitter power on the other channels are turned off. It simply means that for the purpose of calculating co-channel interference we are not interested in the power levels on other channels and those values will not be updated as a result of introducing the new call. The power levels calculated by the ILP only affect transmissions on the channel assigned to the new call. Constraint (7) is based on equation (2), and specifies that the desired signal level for communication using channel *l*, in a cell *i* must be at least $\gamma_o$ times the total co-channel interference plus the thermal noise power in the cell.

## 4. RESULTS

### 4.1. Simulation Model

In our experiments, we assume the traffic model follows the blocked-calls-cleared queuing discipline. An incoming call is served immediately if a free channel with an appropriate power level can found, otherwise the new call is blocked and not queued. Each channel can service at most one call in a cell and channel selection is based on co-channel interference only. We assume that call arrival follows a Poisson process with mean arrival rate of $\lambda$ calls/hour, and call duration is exponentially distributed with mean *x*. Inter-arrival time follows a negative exponential distribution with mean *x*. We consider a non-uniform traffic distribution (where each cell may have a different call arrival rate). The traffic load for each cell is computed as the product of the mean arrival rate and the mean call duration. For experimental purposes, we have used a topological model consisting of a group of 49 hexagonal cells that form a parallelogram shape (taken from [25]), with the traffic pattern shown in Fig. 2. The mean call arrival rate per hour, as shown as an entry in each cell, is given under *normal load* condition.

The network has a total of 70 channels, which are divided into fixed and dynamic channels according to the ratio FC:DC, where |FC ∪ DC| = 70. A channel serves one call at most.

Incoming calls at each cell may be served by any of the free channels from the set FC ∪ DC. Reassignment of existing calls to a new channel is not allowed.

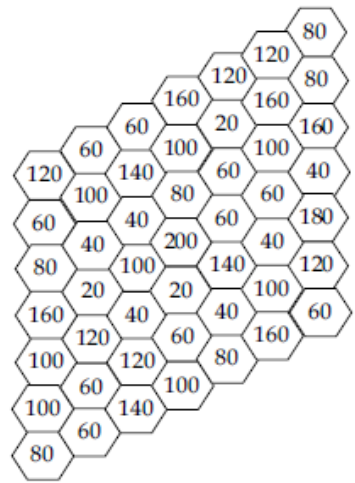

Fig. 2. Non Uniform traffic distribution Pattern

### 4.2. Analysis of Results

We have simulated our networks using three representative ratios of fixed and dynamic channels, FC:DC - 21:49, 35:35 and 49:21, as in [21]. The *blocking probability* for new incoming calls is defined as the ratio between the number of blocked calls and the total number of call arrivals in the system. In order to study the effectiveness of our approach, we have compared the blocking probability for conventional techniques (without power control) with that which can be achieved using our proposed approach with power control (PC). For the conventional systems, we investigated two schemes; one based on fixed power levels (FP) and the second based on the concept of reuse distance (RD) [21]. For these simulations, we set CIR=2 and reuse distance=3. These two parameters can be considered "equivalent" in the sense that in both cases a single channel can accommodate approximately 8-9 calls simultaneously in the entire network. Figures 3 and 4 show the results for the case when FC:DC = 21:49 and 49:21 respectively. Both figures clearly indicate that significant improvements can be obtained using the proposed approach. The results for the FC:DC = 35:35 case, showed a similar pattern.

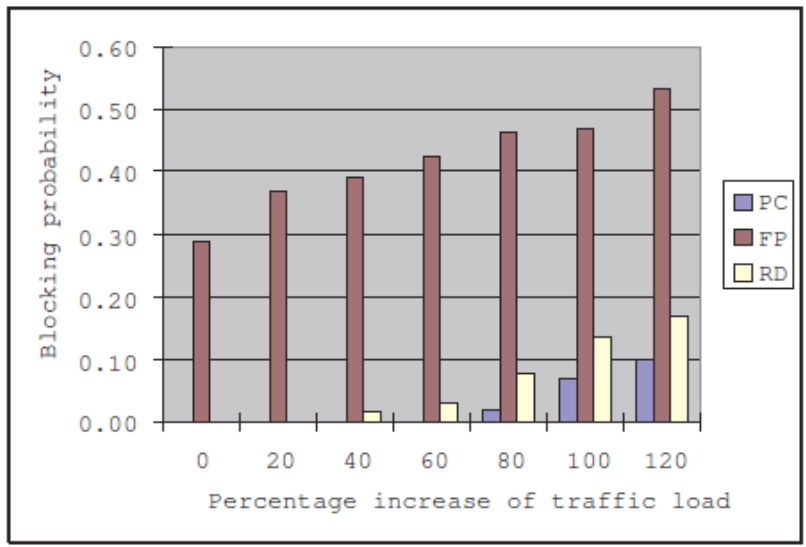

Fig. 3. Blocking probabilities vs. load for ratio 21:49 with CIR = 2.

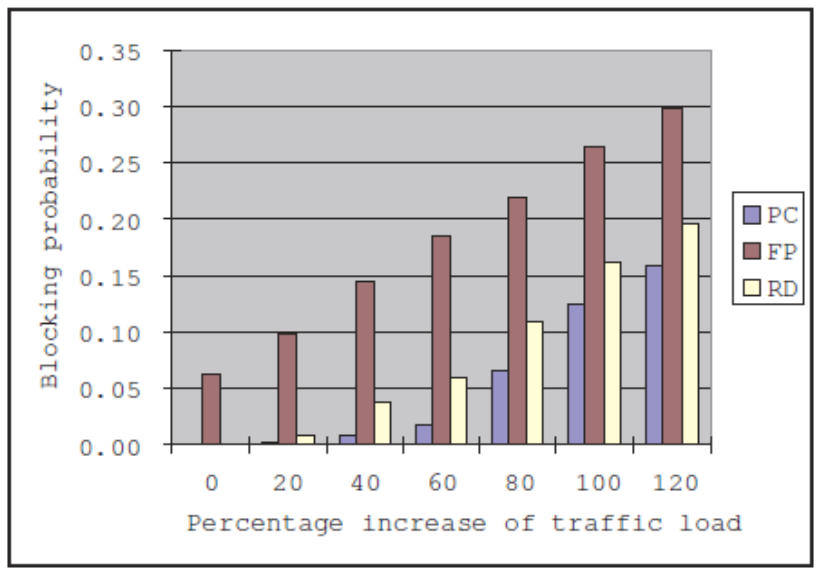

Fig. 4. Blocking probabilities vs. load for ratio 49:21 with CIR = 2.

Fig. 5 compares the relative performance of the different ratios of FC:DC, using the proposed ILP for CIR=2. As shown in the figure, blocking probability increases with traffic, which is expected. Also, 21:49 ratio consistently gives the best performance, followed by 35:35 and 49:21. This is because FC:DC=21:49 offers the greatest flexibility in terms of allocating the channels.

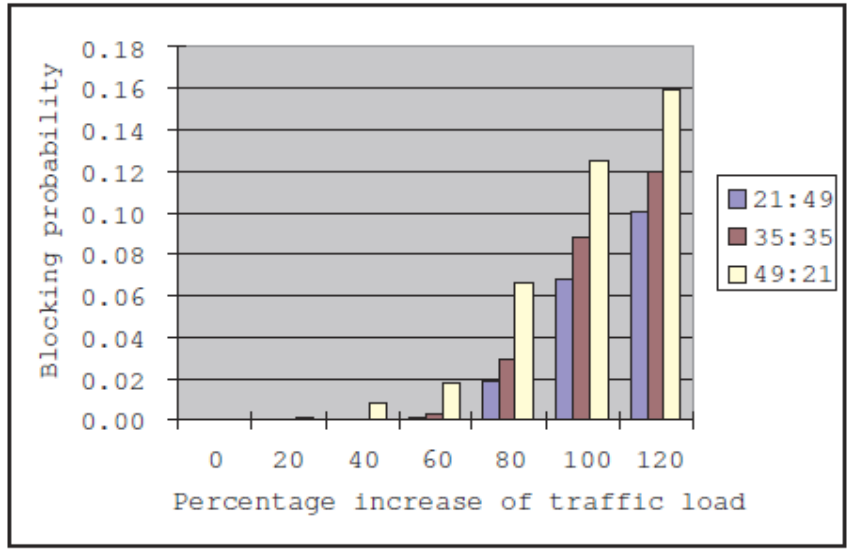

Fig. 5. Comparison of different FC:DC ratios with CIR = 2

Finally, we study how the blocking probability is affected by using different values of CIR. The higher the value of CIR, the better is the QoS for the communication. However, this increase in quality comes at a price. In order to accommodate the higher CIR value, two calls using the same channel must be separated by a larger distance. This results in lower channel utilization and a corresponding increase in blocking probability, as shown in Fig. 6. As mentioned earlier, CIR = 2 is approximately equivalent to a reuse distance of 3, in terms of signal quality.

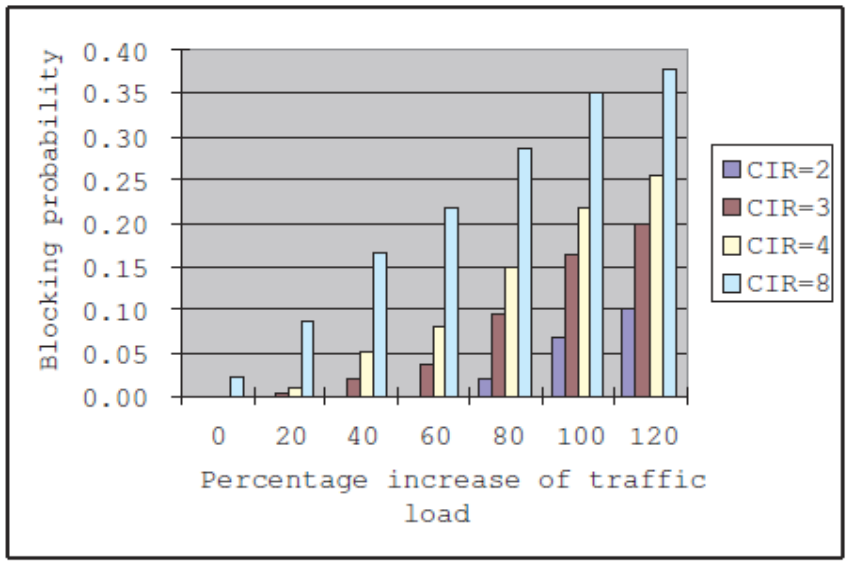

Fig. 6. Effect of CIR on blocking probability for ratio 21:49.

## 5. CONCLUSIONS AND FUTURE RESEARCH DIRECTIONS

In this paper we have proposed an integer linear program (ILP) formulation for combined channel assignment and power control in wireless cellular networks. Our approach not only selects an available channel for a new incoming call, but also determines the appropriate transmission power level for all calls using the selected channel. The goal is to select a channel such that the total power consumption for all calls (including the new call) using that channel is minimized, while ensuring that the CIR of all existing calls, as well as that of the new call are maintained above a specified threshold. Therefore, unlike many previous approaches, we guarantee that no ongoing calls will be dropped due to the introduction of the new call. Experimental results show that our approach leads to a significant reduction in blocking probability, compared to techniques both CIR based (without power control) and reuse distance based channel allocation techniques. We are currently enhancing our ILP by adding constraints to handle co-site and adjacent channel constraints in addition to the co-channel constraints.

## ACKNOWLEDGEMENTS

A. Jaekel has been supported by discovery grants from the Natural Sciences and Engineering Research Council of Canada.

International Journal of Computer Networks & Communications (IJCNC) Vol.3, No.2, March 2011

[28] M. Dosaranian-Moghadam, H. Bakhshi, and G. Dadashzadeh. "Reverse Link Performance of DS-CDMA Cellular Systems through Closed-Loop Power Control and Beamforming in 2D Urban Environment", International Journal of Computer Networks & Communications (IJCNC), Vol.2, No.6, November 2010 DOI : 10.5121/ijcnc.2010.2610 136.

**Authors**

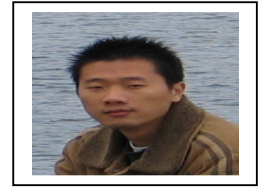

Xin Wu received the B.Eng. degree in software engineering from National University of Defense Technology, Changsha, China, in 1997, and his M.Sc. degree in computer science from University of Windsor, Windsor, Canada, in 2009. His research interests cover wireless mobile computing, wireless sensor networks, network optimization, and distribution systems.

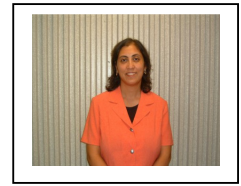

Arunita Jaekel obtained her B. Engg. degree in electronics and telecommunications engineering from Jadavpur University, India. She received her M. A. Sc. and Ph.D. degree in electrical engineering from University of Windsor, Canada. She is currently a professor in the School of Computer Science at University of Windsor. Her research interests include optical networks, survivable topology design and wireless sensor networks.

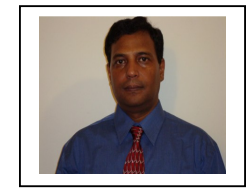

Ataul Bari obtained his PH.D. degree in Computer Science from the University of Windsor in 2010. He obtained his BE degree in Mechanical Engineering from the University of Rajshahi (Engineering College), Bangladesh. He received his BCS(H) and M. Sc. degree in Computer Science from the University of Windsor, Canada. His research interests include wireless sensor networks, cellular networks, optical networks and bioinformatics.